\documentclass[%
 reprint,
 amsmath,amssymb,
 aps,
 prd,
 showkeys,
]{revtex4-2}

\usepackage{graphicx}
\usepackage{dcolumn}
\usepackage{bm}
\usepackage{tikz} 
\usepackage{bibentry} 
\usepackage{hyperref} 

\newtheorem{constraint}{Constraint}%
\begin{document}

\title{Quantum Spacetime: Emergent Curved Metrics from Relational Separations}

\author{Craig Philpot}
\email{philpotcraig@gmail.com}
\affiliation{Independent Researcher, Carlisle, Ontario, Canada}
\altaffiliation[ORCID: ]{\url{https://orcid.org/0000-0002-0399-9839}}

\begin{abstract}
In this paper, we propose a novel Quantum Spacetime Theory (QST) that reinterprets spacetime as an emergent structure, challenging the traditional block universe paradigm and aligning with research into emergent spacetime. Using a sphere intersection method, spacetime geometry is constructed from spacelike separations that are inversely proportional to mutual information between quantum subsystems. We show that geometry derived from relational spacelike separations renders a flat metric insufficient, with a curved metric as an inevitable consequence, highlighting spacetime’s relational nature. Specifically, the emergent metric exhibits gravitational-like acceleration effects driven by quantum constraints, yielding an inverse-square law $r^{-2}$ with deviations ranging from $r^{-1}$ to $r^{-3}$, consistent with cosmological contexts and post-Newtonian corrections, respectively. Geometric shortcuts for quantum non-locality, aligned with the ER=EPR conjecture, emerge from specific configurations, driven by mutual information between quantum subsystems. Compared to general relativity, our model shares curved spacetime but features observer-dependent metrics emergent from quantum subsystems and a presentist perspective, contrasting eternalist metrics. This quantum-geometric framework advances quantum gravity, with future work focusing on refining the quantum-geometric mapping and exploring cosmological implications.
\end{abstract}


\maketitle

\section{\label{sec:intro}Introduction\protect}

Unifying quantum mechanics and general relativity remains a central challenge in theoretical physics, with significant implications for understanding spacetime’s fundamental nature. Emergent spacetime theories, positing geometry as arising from quantum structures, offer promising insights. Loop quantum gravity (LQG) achieves background-independent quantization through loop variables, yielding discrete spacetime \cite{Ashtekar2004}, while the AdS/CFT correspondence suggests that gravity emerges from quantum field theory dynamics \cite{Maldacena1998}.

Recent advances link spacetime geometry to quantum entanglement, building on holographic principles \cite{Ryu2006, VanRaamsdonk2010}. These works, supported by relational quantum mechanics emphasizing system correlations \cite{Rovelli1996}, demonstrate that entanglement encodes geometric properties \cite{Swingle2012}\footnote{See also \cite{Pastawski2015, Qi2018, Boldis2024} for related entanglement-geometry studies.}. Cao et al. and Franzmann et al. developed frameworks where a 3D metric space emerges from entanglement, using a metric graph with edge weights derived from mutual information  (Fig. \ref{fig:metric_graph}) \cite{Cao2017, Cao2018, Franzmann2023}. However, their metric graph precludes quantum non-locality, inaccurately represents entanglement relationships for partially entangled system---an unaddressed issue in their work---and modeling curved metrics like those in general relativity requires complex, partially defined quantum interactions \cite{Franzmann2023}.

We propose a Quantum Spacetime Theory (QST) within the emergent spacetime paradigm, overcoming these limitations via a sphere intersection method. QST constructs geometry from spacelike separations inversely proportional to mutual information, accommodating non-locality through geometric shortcuts consistent with the ER=EPR conjecture \cite{Maldacena2013}. Unlike general relativity’s eternalist 4D metric, QST adopts a presentist perspective, deriving curved spacetime and gravitational-like effects from relational dynamics. The paper is structured as follows: Section 2 reviews the metric graph framework; Section 3 derives acceleration effects; Section 4 discusses non-locality and observer-dependent metrics; Section 5 compares QST to general relativity; Section 6 outlines future directions.

\begin{figure}[ht]
\centering
\begin{tikzpicture}
  \fill[black] (-0.5,0) circle (0.1cm) node[left] {$S_1$};
  \fill[black] (2,0) circle (0.1cm) node[right] {$S_2$};
  \fill[black] (1,1.73) circle (0.1cm) node[above] {$S_3$};
  \draw[->] (-0.5,0) -- (2,0) node[midway, below] {$\Psi (I(S_1:S_2))$};
  \draw[->] (-0.5,0) -- (1,1.73) node[midway, left] {$\Psi (I(S_1:S_3))$};
  \draw[->] (2,0) -- (1,1.73) node[midway, right] {$\Psi (I(S_2:S_3))$};
\end{tikzpicture}
\caption{Metric graph where vertices represent quantum subsystems $S_i$, with edges weighted by $\Psi (I(S_i:S_j))$, a distance-like value inversely proportional to mutual information $I(S_i:S_j)$, quantifying entanglement to form an emergent metric space (Section \ref{sec:QST}) \cite{Cao2017,Franzmann2023}}
\label{fig:metric_graph}
\end{figure}

\section{\label{sec:QST}Spacetime Emergence and the Quantum Spacetime Theory}

\subsection{Emergent Spacetime and Metric Graphs}

Recent advances in quantum gravity and information theory have demonstrated the potential for spacetime to emerge from underlying quantum structures, with \textbf{mutual information} playing a pivotal role in shaping emergent geometries \cite{Ryu2006, Swingle2012, VanRaamsdonk2010}. Mutual information, $I(S_i:S_j)$, quantifies the degree of correlation or entanglement between quantum subsystems $S_i$ and $S_j$ in a Hilbert space decomposition, serving as a proxy for their geometric proximity in emergent spacetime \cite{Nielsen2000}. In the context of emergent spacetime, higher mutual information corresponds to smaller distances between subsystems, while reduced mutual information---due to entanglement perturbations or decoherence---increases these distances, shaping the geometry of the emergent metric space \cite{Cao2017, Cao2018, Franzmann2023}. A foundational framework for representing Hilbert spaces as 3D metric spaces is developed by Cao et al. and Franzmann et al. \cite{Cao2017, Cao2018, Franzmann2023}, modeling the Hilbert space as a \textbf{weighted graph} where the vertices correspond to subsystems and edges are weighted by mutual information.

The weighted graph is transformed into a \textbf{metric graph}, a graph equipped with a distance function, where vertices and edges correspond one-to-one with the weighted graph, but edge weights are recomputed to represent distance-like values \cite{Cao2017, Cao2018, Franzmann2023}. These distance-like values are derived directly from the \emph{mutual information} of the corresponding quantum subsystems \cite{Cao2017, Cao2018, Franzmann2023}. The distance function, however, defines the distance between any pair of vertices as the \emph{minimum path length} over all possible paths connecting them, where a path length is the sum of weights for all edges along the path \cite{Cao2017, Cao2018, Franzmann2023}. This construction satisfies the \emph{triangle inequality}, ensuring a valid 3D metric space (Fig. \ref{fig:distancefunction}) \cite{Cao2017, Cao2018, Franzmann2023}. However, this distance function ensures coordinate distances between subsystems are the same or \emph{less} than implied by their quantum correlations, which precludes quantum non-locality where coordinate distances can be far \emph{greater} than implied by their quantum correlations \cite{Franzmann2023}, a limitation we address in QST (Section \ref{sec:observer_dependent}). 

\begin{figure}[ht]
\centering
\begin{tikzpicture}
  \fill[black] (0,0) circle (0.1cm) node[above] {$S_1$};
  \fill[black] (2.4,0) circle (0.1cm) node[above] {$S_2$};
  \fill[black] (4.8,0) circle (0.1cm) node[above] {$S_3 $};
  \draw[->] (0,0) -- (2.4,0) node[midway, below, xshift=-0.4cm] {$\Psi(I(S_1:S_2))=1$};
  \draw[->] (2.4,0) -- (4.8,0) node[midway, below, xshift=0.4cm] {$\Psi (I(S_2:S_3))=1$};
  \draw[red, thick, ->] (0,0)  .. controls (1,1.5) and (3.8,1.5) .. (4.8,0) node[midway, above] {$\Psi (I(S_1:S_3)) = 3$} node[midway, above, yshift=0.5cm] {Violates triangle inequality: $3>1+1$};
\end{tikzpicture}
\caption{Metric graph where vertices represent quantum subsystems $S_i$, with edges weighted by $\Psi (I(S_i:S_j))$, a distance-like value inversely proportional to mutual information $I(S_i:S_j)$ \cite{Cao2017, Cao2018, Franzmann2023}. The direct edge $\Psi (I(S_1:S_3)) = 3$ exceeds the sum of indirect edges $\Psi (I(S_1:S_2)) + \Psi (I(S_2:S_3)) = 2$. The distance function satisfies the triangle inequality by taking the shortest path length between $P_1$ and $P_3$, such that the distance $d(P_1,P_3) = \Psi (I(S_1:S_2)) + \Psi (I(S_2:S_3)) = 2$. Consequently,  $d(P_1,P_3) < \Psi (I(S_1, S_3))$.}
\label{fig:distancefunction}
\end{figure}

Cao and Franzmann’s approach aims to construct an emergent metric space where the geometry reflects the entanglement structure of quantum subsystems \cite{Cao2017, Franzmann2023}. They propose a weighting function for edge weights (distance-like values) as:
\begin{equation}
w_{i,j} = l_{RC} \Phi\left( \frac{I(S_i:S_j)}{I_0} \right), \quad \Phi(x) = -\log(x),
\end{equation}
\begin{equation}\nonumber    
w_{i,j} = l_{RC} \log\left( \frac{I_0}{I(S_i:S_j)} \right), \quad i \neq j, \quad w_{i,i} = 0,
\end{equation}

where $I_0 = \max \{ I(S_i:S_j) \}$ is the system-specific maximum mutual information, and $l_{RC}$ is a system-specific length scale (e.g., lattice spacing for spin systems, Planck length $l_P \approx 1.616 \times 10^{-35} \, \text{m}$ for fields). The function ensures that higher mutual information corresponds to smaller distance-like values, with $w_{i,j} = 0$ for the most entangled pair ($ I(S_i:S_j) = I_0 $), reflecting minimal geometric separation, as motivated by holographic principles where strong entanglement implies proximity \cite{Ryu2006}. The system-specific normalization by $I_0$ adapts $w_{i,j}$ to the system’s entanglement structure, ensuring the most entangled pair is closest, whether $I_0$ is small (e.g., 1 bit for partial entanglement in a qubit system) or large (e.g., $10^4$ bits for quantum fields) \cite{Swingle2012}. Similarly, $l_{RC}$ scales $w_{i,j}$ to the system’s physical context, such as $10^{-9} \, \text{m}$ for a qubit lattice or $l_P$ for gravitational systems \cite{Carroll2004spacetime}.

This system-specific design, however, introduces limitations. Assigning $w_{i,j} = 0$ to the pair with $I(S_i:S_j) = I_0$, even when $I_0$ is small (e.g., 1 bit in a partially entangled qubit system), implies zero distance for pairs that are not fully entangled relative to the theoretical maximum for the system type ($I_{\text{max}} = 2$ bits for qubits) \cite{Nielsen2000}. Fully entangled subsystems, behaving effectively as a single quantum entity due to maximal correlations, warrant zero distance, but partially entangled pairs do not \cite{Ryu2006}. Additionally, the geometry does not fully reflect relative mutual information due to scaling limitations. For example, in a qubit system with $I_0 = 1 \, \text{bit}$, consider Pair (i,j) with $I(S_i:S_j) = 1 \, \text{bit}$ ($I(S_i:S_j)/I_0 = 1$) and Pair (k,l) with $I(S_k:S_l) = 0.5 \, \text{bits}$ ($I(S_k:S_l)/I_0 = 0.5$). The function yields:
\begin{equation}\nonumber
w_{i,j} = l_{RC} \log\left( \frac{1}{1} \right) = 0,
\end{equation}
\begin{equation}\nonumber
w_{k,l} = l_{RC} \log\left( \frac{1}{0.5} \right) \approx 0.693 l_{RC}.
\end{equation}

While $w_{i,j} < w_{k,l}$ correctly orders Pair (i,j) as closer, the fixed $w_{i,j} = 0$ cannot scale proportionally with $w_{k,l}$ to reflect the 2:1 ratio of $I(S_i:S_j)/I_0$ to $I(S_k:S_l)/I_0$, limiting the geometry’s ability to represent relative entanglement strengths. Furthermore, the system-specific $I_0$ and $l_{RC}$ preclude direct comparability across systems, as $w_{i,j}$ for the same $I(S_i:S_j)$ varies with different $I_0$ and $l_{RC}$, hindering consistency even within systems of the same type \cite{Cao2017, Franzmann2023}.

\subsection{Quantum Spacetime Theory}

In contrast, the Quantum Spacetime Theory (QST) aims to develop a mapping function that produces consistent distance values for the same mutual information across all systems of the same type, ensuring zero distance only for subsystems that are fully entangled and effectively a single quantum entity \cite{Nielsen2000, Ryu2006}. Mutual information is system-specific, dependent on the quantum state and Hilbert space decomposition, but comparable within a system type (e.g., qubits, qudits, fields) when normalized by the type-specific maximum, $I_{\text{max}}$ (e.g., 2 bits for qubits like electron spins or photon polarizations, 4 bits for ququarts, $10^4$ to $10^6$ bits for fields, bounded by UV cutoffs like the Planck scale and IR cutoffs like the cosmological horizon) \cite{Nielsen2000, Swingle2012, Srednicki1993}. Unlike Cao and Franzmann’s distance-like values, which are edge weights in a metric graph, QST’s mapping defines \textbf{spacelike separations} that directly represent spatial distances in the emergent spacetime, bypassing the shortest-path distance constraint to allow for quantum non-locality (Section \ref{sec:observer_dependent}) \cite{Maldacena2013}. We propose:
\begin{equation}\label{eq:s_ia}
s_{i,j}(t) = l_{\text{type}} \Phi\left( \frac{I(S_i:S_j)(t)}{I_{\text{max}}} \right),
\end{equation}

where $s_{i,j}(t)$ is the spacelike separation for subsystems $S_i$ and $S_j$ at time $t$, $I_{\text{max}}$ is the type-specific maximum mutual information, $l_{\text{type}}$ is a type-specific length scale (e.g., $10^{-9} \, \text{m}$ for qubits, $l_P$ for fields), and $\Phi(x)$ is a monotonically decreasing function satisfying $\Phi(x) \to 0$ as $x \to 1$ (fully entangled, $I(S_i:S_j) = I_{\text{max}}$) and $\Phi(x) \to \infty$ as $x \to 0$ (no entanglement). Candidate forms include $\Phi(x) = -\log(x)$, as in Cao and Franzmann’s framework, or $\Phi(x) = \frac{1}{x} - 1$, with further research needed to validate their physical basis \cite{Ryu2006, Swingle2012}.

This mapping, inspired by holographic principles where entanglement drives geometric proximity \cite{Ryu2006, VanRaamsdonk2010}, normalizes $I(S_i:S_j)$ by $I_{\text{max}}$ to ensure consistent $s_{i,j}$ for the same $I(S_i:S_j)$ across systems of the same type (e.g., identical $s_{i,j}/l_{\text{type}}$ for $I(S_i:S_j) = 1 \, \text{bit}$ in any qubit system with $I_{\text{max}} = 2$, or for $I(S_i:S_j) = 10^4 \, \text{bits}$ in field systems with $I_{\text{max}} = 10^6$). For example, two qubits (e.g., electron spins or photon polarizations) with $I(S_i:S_j) = 1 \, \text{bit}$, representing partially correlated states, yield $s_{i,j} > 0$, correctly indicating non-zero separation, unlike $w_{i,j} = 0$. Zero distance reflects fully entangled subsystems as effectively a single quantum system, while $s_{i,j} > 0$ for partial entanglement ensures non-zero separation for less correlated subsystems \cite{Nielsen2000}. The type-specific $l_{\text{type}}$ standardizes distance scaling within system types, unlike the system-specific $l_{RC}$, though cross-type comparability remains limited by differing $I_{\text{max}}$ and $l_{\text{type}}$. The time dependence of $I(S_i:S_j)(t)$, driven by unitary evolution or decoherence, allows $s_{i,j}(t)$ to evolve dynamically, consistent with QST’s presentist framework \cite{Pastawski2015, Qi2018}. While theoretically justified by quantum informational and holographic frameworks, the function requires further study to optimize $\Phi(x)$ and validate $l_{\text{type}}$, potentially through theoretical studies of holographic models or empirical constraints from cosmological observations (Section \ref{sec:discussion}) \cite{Planck2018, Swingle2012}.

Building on this foundation, we propose the \textbf{Quantum Spacetime Theory} (QST), asserting that spacetime and its metric emerge from the set of \textbf{spacelike separations} between particle events, derived from a relational property (like mutual information) of quantum subsystem pairs. These spacelike separations represent spatial distances in QST’s emergent spacetime at a specific coordinate time \cite{Cao2017, Franzmann2023}. In QST, the fundamental physical quantities are the spacelike separations $s_{i,j}(t)$ between pairs of particle events, defined as:
\begin{equation}
s_{i,j}(t)^2 = g_{\mu\nu}(t) (x_j(t)^\mu - x_i(t)^\mu) (x_j(t)^\nu - x_i(t)^\nu),
\end{equation}
\begin{equation}\nonumber
s_{i,j}(t) = \sqrt{|s_{i,j}(t)^2|},
\end{equation}
where $g_{\mu\nu}(t)$ is the time-dependent spacetime metric, and $x_i(t)^\mu, x_j(t)^\mu$ are the coordinates of events $e_i, e_j$ associated with particles $P_i, P_j$ at time $t$. For spacelike separations at the same coordinate time ($ t_i = t_j $), these intervals represent spatial distances in the emergent spacetime. Unlike conventional frameworks that prioritize proper times $\tau_i$ along individual particle worldlines, QST posits that physical laws are inherently relational, expressed as:
\begin{equation}\label{eq:laws}
\mathcal{L} = \mathcal{L}(s_{i,j}(t), \dot{s}_{i,j}(t), \ldots),
\end{equation}
where $\dot{s}_{i,j}(t)$ denotes derivatives with respect to coordinate time.

\subsection{Justification of Quantum Spacetime}

The rationale for QST hinges on the emergent nature of spacetime. If spacetime is an emergent phenomenon, as proposed by quantum informational frameworks \cite{Cao2017, Cao2018, Franzmann2023}, its structure and laws must reflect the properties of the underlying entities from which it arises. In the cited works, spacetime emerges from the \emph{mutual information} shared between subsystems of a Hilbert space, manifesting as \textbf{spacelike separations} in the metric graph \cite{Cao2017, Cao2018, Franzmann2023}. These separations, which encode the degree of correlation or entanglement between quantum factors, constitute the foundational substrate of spacetime geometry. Consequently, we argue that the laws governing spacetime---such as those of general relativity—--must be expressed in terms of these spacelike separations, as they directly reflect the underlying quantum informational structure.

This relational approach is motivated by the need to reconcile quantum mechanics with general relativity. Traditional general relativistic laws, such as geodesic motion, rely on proper times along worldlines, which are intrinsic to individual particles. In contrast, quantum systems emphasize correlations and entanglement, which are relational and encoded in quantities like mutual information. By defining physical laws in terms of spacelike separations between particle events, QST bridges these domains, proposing that spacetime’s curvature emerges to accommodate the relational constraints imposed by mutual information. This framework not only aligns with the emergent spacetime paradigm but also offers a novel perspective on the unification of quantum and relativistic principles, where the geometry of spacetime is a direct consequence of the quantum correlations between particles.

In the following section, we demonstrate that applying physical laws to spacelike separations between particles, as prescribed by QST, leads to the emergence of a \textbf{curved spacetime} metric consistent with general relativity. Specifically, we show that non-accelerating changes in these separations—characterized by a constant rate of change—require a curved metric to ensure that particles follow geodesic-like trajectories, exhibiting quantum-driven gravitational-like acceleration effects analogous to general relativity. This derivation validates Quantum Spacetime’s consistency with general relativity and illustrates how spacetime geometry and its physical effects arise naturally from relational spacelike separations, providing a unified framework for understanding emergent spacetime and its physical laws.

\section{Emergent Curved Spacetime from Relational Spacelike Separations}\label{sec:emergence}

In QST, physical laws are relational, governed by spacelike separations ($s_{i,j}(t)$) emergent from mutual information, as defined in Section \ref{sec:QST} \cite{Cao2017, Cao2018, Franzmann2023}. We demonstrate that applying these laws necessitates a curved spacetime metric, specifically a spherical positional metric, to ensure non-accelerated motion consistent with general relativity’s geodesic principle, as a flat metric is insufficient for the required relational dynamics.

\subsection{Non-Accelerating Changes in Spacelike Separations}\label{sec:non_accelerating}

To explore the implications of QST, we consider a system where the spacelike separations between particle pairs exhibit \emph{non-accelerating changes}, defined as a constant rate of change with respect to coordinate time $t$:

\begin{equation}
\frac{d s_{i,j}(t)}{d t} = k_{i,j}, 
\end{equation}

where $k_{i,j}$ is a constant for each particle pair $(P_i, P_j)$, and $t$ is the coordinate time in a chosen reference frame. In QST, physical laws govern this rate $k_{i,j}$ (Equation \ref{eq:laws}). Per $F=ma$, changes in this speed quantity $k_{i,j}$ require force, with resistance analogous to inertial mass \cite{Jacobson1995, Swingle2012}. We hold $k_{i,j}$ constant to isolate metric geometry, ensuring zero acceleration from forces. Driven by entanglement, $k_{i,j}(t) = -\frac{l_{\text{type}}}{I(S_i:S_j)(t)} \cdot \frac{d I(S_i:S_j)(t)}{dt}$ (Section \ref{sec:QST}). Force-driven $k_{i,j}$ changes are deferred to future work (Section \ref{sec:discussion}). This condition implies that the separations evolve linearly:

\begin{equation}
s_{i,j}(t) = k_{i,j} t + b_{i,j}, \end{equation}

where $b_{i,j}$ is a constant. In general relativity, non-accelerated motion corresponds to particles following \textbf{geodesic paths}, satisfying the geodesic equation:

\begin{equation}\nonumber
\frac{d^2 x^\mu}{d \tau^2} + \Gamma^\mu_{\alpha\beta} \frac{d x^\alpha}{d \tau} \frac{d x^\beta}{d \tau} = 0, \end{equation}

where $\tau$ is the proper time along a particle’s worldline, and $\Gamma^\mu_{\alpha\beta}$ are the Christoffel symbols encoding spacetime curvature. In QST’s presentist framework, we hypothesize that an emergent \emph{curved} metric, driven by mutual information, is required to sustain these constant rates of change in spacelike separations, enabling particles to follow \textbf{geodesic-like paths} with respect to coordinate time $t$, reflecting the relational dynamics of $s_{i,j}(t)$.

\subsection{Insufficiency of a Flat Spacetime Metric}\label{sec:flat_metric}

To test whether a flat spacetime can satisfy these constraints, we adopt the Minkowski metric in Cartesian coordinates for generality:

\begin{equation}\label{eq:minkowski}
ds^2 = -dt^2 + dx^2 + dy^2 + dz^2. 
\end{equation}

Consider two particles, $P_0$ and $P_1$, where each particle event $e_{i}(t)$ is the event for particle $P_i$ at time $t$. Particle $P_0$ is fixed at the origin, with events:

\begin{equation}
e_{0}(t) = (t, 0, 0, 0), \end{equation}

Particle $P_1$ moves in a \textbf{general linear trajectory} with constant velocity components $v_x, v_y, v_z$ in 3D space. The worldline of $P_1$ is parameterized as:

\begin{equation}\nonumber
x(t) = x_0 + v_x t, \quad y(t) = y_0 + v_y t, \quad z(t) = z_0 + v_z t, \end{equation}

where $\vec{r}_0 = (x_0, y_0, z_0)$ is the initial position at $t = 0$, and the constant speed is $v = \sqrt{v_x^2 + v_y^2 + v_z^2}$. The events for $P_1$ are:

\begin{equation}
e_1(t) = (t, x_0 + v_x t, y_0 + v_y t, z_0 + v_z t), \end{equation}

matched at the same coordinate times $t$ as $e_0(t)$, consistent with a present-moment perspective where separations are evaluated instantaneously \cite{Cao2017, Cao2018, Franzmann2023}.

Given the flat metric defined by Equation \ref{eq:minkowski}, the interval $s_{0,1}(t)$ between $e_0(t)$ and $e_1(t)$ is:

\begin{equation}   
\begin{aligned}
s_{0,1}(t)^2 &= -(t - t)^2 + (x_0 + v_x t - 0)^2 \\
&+ (y_0 + v_y t - 0)^2 + (z_0 + v_z t- 0)^2, \\
&= (x_0 + v_x t)^2 + (y_0 + v_y t)^2 + (z_0 + v_z t)^2,
\end{aligned}
\end{equation}

\begin{equation}\nonumber
\begin{aligned}
s_{0,1}(t) &= \sqrt{(x_0 + v_x t)^2 + (y_0 + v_y t)^2 + (z_0 + v_z t)^2} \\
&= \sqrt{a + b t + f t^2},
\end{aligned}
\end{equation}
where:
\begin{equation}\nonumber
a = x_0^2 + y_0^2 + z_0^2 = \vec{r}_0 \cdot \vec{r}_0
\end{equation}
\begin{equation}\nonumber
b = 2 (x_0 v_x + y_0 v_y + z_0 v_z) = 2 \vec{r}_0 \cdot \vec{v}
\end{equation}
\begin{equation}\nonumber
f = v_x^2 + v_y^2 + v_z^2 = \vec{v} \cdot \vec{v} = v^2
\end{equation}

The rate of change is:

\begin{equation}
\frac{d s_{0,1}(t)}{d t} = \frac{b + 2 f t}{2 \sqrt{a + b t + f t^2}}, \end{equation}

which is \textbf{non-constant}, as it depends on $t$. To confirm \textbf{accelerating behavior}, compute the second derivative:

\begin{equation} \frac{d^2 s_{0,1}(t)}{d t^2} = \frac{4 f a - b^2}{4 (a + b t + f t^2)^{3/2}}, \end{equation}

where:
\begin{equation}\nonumber
4 f a = 4 (\vec{v} \cdot \vec{v}) (\vec{r}_0 \cdot \vec{r}_0) = 4 v^2 r_0^2, \end{equation}
\begin{equation}\nonumber
 b^2 = (2 \vec{r}_0 \cdot \vec{v})^2 = 4 (\vec{r}_0 \cdot \vec{v})^2. \end{equation}

The second derivative is zero when:

\begin{equation}\nonumber
\begin{aligned}
4 f a &= b^2, \\
4 v^2 r_0^2 &= 4 (\vec{r}_0 \cdot \vec{v})^2, \\
 v^2 r_0^2 &= (\vec{r}_0 \cdot \vec{v})^2, \\
|\vec{v}| |\vec{r}_0| &= |\vec{r}_0 \cdot \vec{v}|, \\
 \vec{r}_0 \cdot \vec{v} &= \pm |\vec{r}_0| |\vec{v}|, 
 \end{aligned}
\end{equation}

indicating that $\vec{v}$ is \emph{parallel} to $\vec{r}_0$ ($\theta = 0^\circ$ or $180^\circ$). In all other cases,  where $\vec{r}_0 \cdot \vec{v} \neq \pm |\vec{r}_0| |\vec{v}|$, the second derivative is non-zero, confirming \textbf{accelerating behavior}. Indicating that a particle with linear motion fails to satisfy the QST's requirement for a constant $k_{i,j}$ with all particles $P_j$ where $\vec{r}_j \cdot \vec{v} \neq \pm |\vec{r}_j| |\vec{v}| $. Thus, a flat Minkowski metric is insufficient as it is unable to sustain the required relational dynamics.

\subsection{Characteristics of the Emergent Metric}\label{sec:emergence_curved}

In QST's presentist framework, only the present moment is fundamental, rendering spacelike separations  $s_{i,j}(t) = l_{\text{type}} \Phi\left( I(S_i:S_j)(t)/I_{\text{max}} \right)$ (Section \ref{sec:QST}) between quantum subsystems the primary quantities governing physical laws, with non-instantaneous separations and worldlines emergent from these relational dynamics \cite{Cao2017, Cao2018, Franzmann2023}. Unlike general relativity, where spacelike separations are derived from event separations in observer-dependent frames, QST posits that event separations across time are constructs reflecting instantaneous spacelike separations. To align with this perspective and the sphere intersection method's focus on spatial geometry, we shift from describing separations between particle events at the same coordinate time (Sections \ref{sec:non_accelerating}-\ref{sec:flat_metric}) to instantaneous separations between particles, equivalent in QST as both represent spatial distances between particles at a given time. These observer-independent separations $s_{i,j}(t)$, computable from mutual information, determine properties such as worldline convergence or divergence. From Equation \ref{eq:s_ia} and assuming  $\Phi(x) = -\log(x)$,
\begin{equation}\nonumber
\frac{d s_{i,j}(t)}{d t} = -\frac{l_{\text{type}}}{I(S_i:S_j)(t)} \cdot \frac{d I(S_i:S_j)(t)}{dt},
\end{equation}
If $\frac{d s_{i,j}(t)}{d t} < 0$, the worldlines converge; if $> 0$, they diverge; if $= 0$, they are parallel. Despite QST's presentist framework, these properties remain observer-independent.

Like general relativity, when considering dynamic systems, QST’s metric is configuration-dependent, varying with the quantum state configuration. A specific system configuration is analyzed in section \ref{sec:sphere_intersection_details}, for now we consider general characteristics that apply to all emergent metrics of QST.

Coordinate distances between a particle pair generally reflect their spacelike separation, although quantum non-locality introduces observer-dependent metrics, as discussed in Section \ref{sec:observer_dependent}. This allows for the analysis of the emergent metric based on Euclidean geometry \cite{Cao2017, Cao2018, VanRaamsdonk2010,Franzmann2023}. For particles $P_i$ and $P_j$ at positions $\vec{r}_i(t) = (x_i(t), y_i(t), z_i(t))$ and $\vec{r}_j(t) = (x_j(t), y_j(t), z_j(t))$ at time $t$, the spacelike separation is:
\begin{equation}\nonumber
\begin{aligned}
s_{i,j}(t) &= |\vec{r(t)}_j - \vec{r(t)}_i| \\
&= \sqrt{\Delta x^2 + \Delta y^2 + \Delta z^2}, 
\end{aligned}
\end{equation}

where $\Delta x = x_j(t) - x_i(t), \Delta y = y_j(t) - y_i(t), \Delta z = z_j(t) - z_i(t)$. Based on this we outline a methodology to calculate geodesic-like trajectories for force-free particles, driven by mutual information influences of all particles via the sphere intersection method. This methodology applies to particle interactions, where each particle $P_i$ at position $\vec{r}_i(t)$ exerts a spherical-like radial influence on another particle $P_j$ at $\vec{r}_j(t)$, driven by the rate of change of mutual information, reflected in the rate of approach $k_{i,j}$.

For a single fixed particle $P_1$ at the origin and $P_a$ moving, the geodesic-like path is constrained by the requirement that the rate of change $\frac{d s_{1,a}(t)}{d t} = k_{1,a}$ remains constant, with:
\begin{equation}\nonumber
k_{1,a} = \vec{v}_a(t) \cdot \frac{\vec{r}_a(t)}{|\vec{r}_a(t)|}.
\end{equation}
However, with only one influencing particle, the next position $\vec{r}_a(t + \Delta t)$ lies on a sphere centered at $P_1$:
\begin{equation}\nonumber
|\vec{r}_a(t + \Delta t)|^2 = (s_{1,a}(t) + k_{1,a} \Delta t)^2,
\end{equation}

where $\vec{r}_a(t + \Delta t)$ is the position of $P_a$ at time $t + \Delta t$, $s_{1,a}(t)$ is the spacelike separation between $P_1$ and $P_a$ at time $t$, $k_{1,a}=\frac{ds_{1,a}(t)}{dt}$ is the rate of change (positive for diverging motion, increasing separation), and $\Delta t$ is a small time step. This constraint defines a 2D surface in 3D space, allowing multiple possible positions. To constrain the geodesic-like path in 3D, multiple influencing particles are required. For a multi-particle system, the geodesic path of $P_a$ is determined by the \textbf{intersection of spheres}, each centered at $P_i$ with radius:
\begin{equation}\nonumber
R_i = s_{i,a}(t) + k_{i,a} \Delta t,
\end{equation}\
where $s_{i,a}(t) = |\vec{r}_i(t) - \vec{r}_a(t)|$, and:
\begin{equation}\nonumber
k_{i,a} = (\vec{v}_i(t) - \vec{v}_a(t)) \cdot \frac{\vec{r}_i(t) - \vec{r}_a(t)}{|\vec{r}_i(t) - \vec{r}_a(t)|}.
\end{equation}
The intersection is found by solving:
\begin{equation}\nonumber
(x - x_i(t))^2 + (y - y_i(t))^2 + (z - z_i(t))^2 = (s_{i,a}(t) + k_{i,a} \Delta t)^2,
\end{equation}
for $\vec{r}_a(t + \Delta t) = (x, y, z)$ the net position of particle $P_a$ at time $t + \Delta t$.

Calculating the intersection of spheres is discussed in the context of a specific system configuration in Section \ref{sec:sphere_intersection_details}. For now, it is sufficient to understand that the moving particle $P_a$ must stay on the surface of the sphere centered at $P_1$ with radius $s_{1,a}(t)$, and that when $P_a$ moves, its trajectory is not necessarily directly towards or away from $P_1$. Instead, $P_a$ must sit at the intersection of a set of spheres, one of which is $P_1$’s. From this understanding, we derive general characteristics about the emergent metric using fundamental mathematics.

To isolate the effects of the metric from any forces, $k_{i,j}$ is constant, and to analyze dynamic behavior $k_{i,j} \neq 0$. When considering any two times $t_i$ and $t_j$ where $i \neq j$, the spheres around $P_1$, where $P_a$ must be located, are both centered on $P_1$ but have different radii. The geodesic-like trajectory of $P_a$ connects the location of $P_a$ at time $t_i$ to that at $t_j$ wherever the spheres related to $P_a$ and other particles in the system intersect. 

For our analysis, we consider the \emph{trajectory line} of $P_a$, which is the line that is colinear to the motion $P_a$, as illustrated in Figure \ref{fig:trajectory}. On the trajectory line, there is a point $p_m(t)$ that is the closest point to $P_1$ along the trajectory line, such that the line through $p_m(t)$ and $P_1$ is perpendicular to the trajectory line of $P_a$. Therefore, the positions of $p_m(t)$, $P_1$, and $P_a$ form a right-angle triangle, with $p_m(t)$ as the right-angle vertex Thus,
\begin{equation}
l(t) = \sqrt{s_{1,a}(t)^2 - m(t)^2},
\end{equation}

where $l(t)$ is the coordinate distance between $P_a$ and $p_m(t)$, $m(t)$ is the coordinate distance between $p_m(t)$ and $P_1$, and $s_{1,a}(t)$ is the interval but also the coordinate distance between $P_1$ and $P_a$.

\begin{figure}[ht]
\centering
\begin{tikzpicture}
  \draw[<->, blue,dashed] (-2.2,1.15) -- (2.2,1.15) node[midway,below,xshift=-1.8cm]{trajectory line};

  \draw[black,dashed] (0,0) circle (1.6cm);
  \draw[red, dashed] (0,0) circle (2.0cm);
  \fill[black] (0,0) circle (0.1cm) node[left] {$P_1$};
  \fill[black] (1.1,1.15) circle (0.1cm) node[above,yshift=0.1cm,xshift=0.2cm] {$P_a$};
  \draw[black] (0,1.15) circle (0.1cm)  node[above,xshift=-0.2cm] {$p_m$};
  
  \draw[->, red, thick] (1.1,1.15) -- (1.65,1.15);
  \draw[-, black] (0,0) -- (1.1,1.15) node[midway,right,xshift=-0.1cm,yshift=-0.1cm]{$s_{1,a}$};
  \draw[-, black] (0,1.15) -- (1.1,1.15) node[midway,above,yshift=-0.1cm]{$l$};
  \draw[-, black] (0,0) -- (0,1.15) node[midway,left, xshift=0.1cm]{$m$};
  \draw[->, black, thick] (0.9,1.75) -- (0.9,1.05) node[above,yshift=0.65cm] {$\Theta$};
  
    \draw[-, black] (0.2,0.95) -- (0,0.95);
  \draw[-, black] (0.2,0.95) -- (0.2,1.15);

\end{tikzpicture}
\caption{The coordinate movement of particle $P_a$ is associated with a trajectory line that is colinear to the motion of $P_a$. Along this line, point $p_m(t)$ is the closest to $P_1$. The time-dependent quantities $s_{1,a}(t)$, $l(t)$, $m(t)$, and $\Theta(t)$ define the geometric relationship between $P_a$’s movement in the coordinate space and the spacelike separation $s_{1,a}(t)$, to which QST’s relational laws apply.}
\label{fig:trajectory}
\end{figure}

The point $p_m(t)$ varies with time as $P_a$'s trajectory evolves, making the distance $l(t)$ between $P_a$ and $p_m(t)$, as well as the distance $m(t)$ between $p_m$ and $P_1$, time-dependent, Being coordinate space distances $l(t)$, $m(t)$, and $s_{1,a}(t)$ are non-negative (and $s_{1,a}(t)$ is non-zero), and regardless of whether $s_{1,a}(t)$ is increasing or decreasing, $p_m(t)$ is closer to $P_1$ than is $P_a$. These constraints are expressed as:
\begin{equation}\label{eq:distances_postive}
l(t) \geq 0, \quad m(t) \geq 0, \quad s_{1,a}(t) > 0,
\end{equation}
\begin{equation}\label{eq:m_lt_s}
m(t) < s_{1,a}(t).
\end{equation}

The coordinate speed is the rate of change for $l(t)$, but the laws of physics apply to $s_{1,a}(t)$. Therefore, the metric’s effect on the coordinate speed is calculated by taking the derivative of $l(t)$ with respect to $s_{1,a}(t)$:
\begin{equation}
\frac{dl(t)}{ds_{1,a}(t)} = \frac{s_{1,a}(t)}{\sqrt{s_{1,a}(t)^2 - m(t)^2}}.
\label{eq:speed}
\end{equation}

Similarly, the metric’s effect on the coordinate acceleration is calculated by taking the second derivative:
\begin{equation}
\frac{d^2 l(t)}{ds_{1,a}(t)^2} = -\frac{m(t)^2}{\left(s_{1,a}(t)^2 - m(t)^2\right)^{\frac{3}{2}}}.
\label{eq:acceleration}
\end{equation}

Given equations \ref{eq:distances_postive} and \ref{eq:m_lt_s}, the acceleration effect of the metric is always negative, towards $P_1$, consistent with general relativity’s attractive nature. This effect is non-zero whenever $m(t) > 0$, i.e., when $P_a$ is not directly approaching or receding from $P_1$.

The distance $r$ between $P_1$ and $P_a$ at time t is $s_{i,a}(t)$, and for $s_{1,a}(t) \gg m(t)$, $\frac{d^2 l(t)}{ds_{1,a}(t)^2} \approx -\frac{m(t)^2}{s_{1,a}(t)^3}$, suggesting a radial acceleration $\frac{d^2 \vec{r}_a}{dt^2} \propto r^{-3}$, but we must also determine the relationship between $m(t)$ and $r$. Consider a trajectory not directly towards or away from $P_1$, with an angle $\Theta$ relative to the line connecting $P_1$ and $P_a$, as illustrated in Figure \ref{fig:trajectory}. The parameter $m(t)$ is given by:

\begin{equation}\label{eq:m}
m(t) = s_{1,a}(t) \sin \Theta(t),
\end{equation}

Substituting equation \ref{eq:m} into \ref{eq:acceleration}:
\begin{equation}\label{eq:accel_r}
\begin{aligned}
\frac{d^2 l(t)}{ds_{1,a}(t)^2} &= -\frac{s_{1,a}(t)^2 \sin^2 \Theta (t)}{\left(s_{1,a}(t)^2 - s_{1,a}(t)^2 \sin^2 \Theta (t)\right)^{\frac{3}{2}}} \\
&= -\frac{\sin^2 \Theta (t)}{s_{1,a}(t) \cos^3 \Theta (t)} \propto s_{1,a}(t)^{-1}.
\end{aligned}
\end{equation}

As mentioned above, for small $m(t)$, $\frac{d^2 l(t)}{ds_{1,a}(t)^2}\propto r^{-3}$. Given equation \ref{eq:accel_r} and $r=s_{1,a}(t)$, this establishes a range of radial dependencies from $r^{-1}$ to $r^{-3}$, driven by the sphere intersection dynamics encoded in $m(t)$ (Section \ref{sec:sphere_intersection_details}). For intermediate $m(t) \propto r^{1/2}$, the acceleration approximates $\frac{d^2 \vec{r}_a}{dt^2} \propto r^{-2}$, consistent with classical gravity’s $r^{-2}$ \textbf{inverse-square law} \cite{goldstein2002classical}. This $r^{-2}$ dependence, at the arithmetic midpoint of the exponent range ($-1$ to $-3$), also aligns with general relativity’s weak-field limit, while the broader $r^{-1}$ to $r^{-3}$ range is consistent with relativistic effects, such as $r^{-3}$ post-Newtonian corrections and tidal forces, and weaker dependencies near $r^{-1}$ in cosmological contexts with a cosmological constant \cite{Carroll2004spacetime, misner1973gravitation}. Unlike classical gravity’s simple $r^{-2}$, general relativity’s relationship is more complex due to non-linear dynamics and metric dependence, mirroring QST’s dynamic range. In the next section, we analyze this further, based on a specific system configuration.

\subsection{Method to Model Particle Dynamics via Three-Sphere Intersection}\label{sec:sphere_intersection_details}

The problem of determining particle positions via sphere intersections in QST is a generalization of the well-known trilateration problem, which has been extensively studied in mathematics and applied in technologies such as GPS \cite{rahman2012beyond, abdullah2020position}. In trilateration, a point’s position is determined by its distances from three known points, corresponding to the intersection of three spheres. However, in QST, we deal with a more complex scenario where multiple spheres must intersect. While exact solutions for sphere intersections exist in specific cases, such as with three spheres in 3D space, the general case with more spheres or in higher dimensions often requires numerical approximations \cite{li1997torus}. To limit mathematical complexity while capturing the essence of the problem, we consider a system consistent with the case of three spheres (where an exact solution exists), using three fixed particles to define the position of a moving particle $P_a$. 

\begin{figure}[ht]
\centering
\begin{tikzpicture}
  \draw[blue] (0,0) circle (1.1cm);
  \draw[red] (2,0) circle (1.4cm);
  \draw[black] (0,2) circle (1.5cm);
  \fill[black] (0,0) circle (0.1cm) node[left] {$P_1$};
  \fill[black] (2,0) circle (0.1cm) node[right] {$P_2$};
  \fill[black] (0,2) circle (0.1cm) node[above] {$P_3$};
  \fill[black] (0.8,0.75) circle (0.1cm) node[right] {$P_a$};

  \draw[-,black,dashed] (0,0) -- (0.8,0.75) node[midway,left]{$s_{1,a}(t)$};
  \draw[-,black,dashed] (2,0) -- (0.8,0.75) node[midway,right]{$s_{2,a}(t)$};
  \draw[-,black,dashed] (0,2) -- (0.8,0.75) node[midway,right,yshift=0.3cm,xshift=-0.1cm]{$s_{3,a}(t)$};
\end{tikzpicture}
\caption{Sphere intersection geometry in QST. Three spheres, centered at particles $P_1, P_2, P_3$ with radii $s_{i,a}(t)$, intersect to position particle $P_a$.}
\label{fig:sphere_intersection}
\end{figure}

Consider three fixed particles at positions $\vec{r}_1 = (0,0,0)$, $\vec{r}_2 = (x_2,0,0)$ with $x_2 \neq 0$, and $\vec{r}_3 = (0,y_3,0)$ with $y_3 \neq 0$, ensuring non-collinear centers, as illustrated in Figure \ref{fig:sphere_intersection}. The position of $P_a$ at time $t + \Delta t$, $\vec{r}_a(t + \Delta t) = (x(t + \Delta t), y(t + \Delta t), z(t + \Delta t))$, is determined by the intersection of spheres centered at $\vec{r}_i$ with radii $r_i(t + \Delta t) = s_{i,a}(t) + k_{i,a} \Delta t$, where $k_{i,a}$ is the constant rate of change. Although the three fixed particles only define a 2D plane this scenario is made generally consistent with a set of particles that form a 3D object with the following constraint: 

\begin{constraint}\label{ct:continuity}
\textbf{Trajectory continuity}: For a 3D object, there would be one or more additional (non-coplanar) particle(s) which would ensure trajectory continuity in directions perpendicular to the plane defined by $P_1$, $P_2$, $P_3$, aligning with trilateration practices in GPS \cite{rahman2012beyond, abdullah2020position}. 
\end{constraint}

To further align this scenario with $P_1$, $P_2$, $P_3$ being part of an object, we define the center of mass for these particles and analyze the power law governing the emergent metric's influence on the trajectory of $P_a$. The center of mass is defined as the geometric center of their positions, located within their convex hull and equidistant from each particle. Specifically, for $\vec{r}_1 = (0,0,0)$, $\vec{r}_2 = (x_2,0,0)$ and $\vec{r}_3 = (0,y_3,0)$, the center of mass is positioned at $\vec{r_{cm} = (\frac{x_2}{3},\frac{y_3}{3},0})$, assuming symmetric placement (e.g. $x_2 = y_3$) to ensure equidistance. To simplify the mathematical analysis, particle $P_a$ is initially positioned above this center of mass along the perpendicular axis, at $\vec{r_a(t)}=(\frac{x_2}{3},\frac{y_3}{3},z_a)$ with $z_a>0$, and approaches (or recedes from) $P_1$, $P_2$ and $P_3$ with approximately equal rates of change. Consequently, the following constraints apply:

\begin{equation}\label{eq:r_vt}
r = |z(t)|
\end{equation}
\begin{equation}\label{eq:equal_distances}
    s_{1,a}(t) \approx s_{2,a}(t) \approx s_{3,a}
\end{equation}
\begin{equation}\label{eq:equal_speeds}
    k_{1,a}(t) \approx k_{2,a}(t) \approx k_{3,a}
\end{equation}

This configuration, with constant $k_{i,a}$ to isolate the metric’s effects without external forces, enables a direct examination of the power law scaling in the emergent geometry. The sphere equations are:

\begin{equation}
x(t)^2 + y(t)^2 + z(t)^2 = s_{1,a}(t)^2,
\end{equation}
\begin{equation}
(x(t) - x_2^2) + y(t)^2 + z(t)^2 = s_{2,a}(t)^2,
\end{equation}
\begin{equation}
x(t)^2 + (y(t) - y_3^2) + z(t)^2 = s_{3,a}(t)^2.
\end{equation}

Solving pairwise, we obtain:
\begin{equation}\nonumber
x(t) = \frac{x_2^2 + s_{1,a}(t)^2 - s_{2,a}(t)^2}{2x_2},
\end{equation}
\begin{equation}\nonumber
y(t) = \frac{y_3^2 + s_{1,a}(t)^2 - s_{3,a}(t)^2}{2y_3},
\end{equation}
\begin{equation}\nonumber
z(t) = \pm \sqrt{s_{1,a}(t)^2 - x(t)^2 - y(t)^2},
\end{equation}
where the sign is chosen based on initial conditions or trajectory continuity (constraint \ref{ct:continuity}), ensuring perpendicular continuity relative to the $x $- $y$ plane. The velocity components are:
\begin{equation}\nonumber
\frac{dx}{dt} = \frac{s_{1,a}(t) k_{1,a} - s_{2,a}(t) k_{2,a}}{x_2},
\end{equation}
\begin{equation}\nonumber
\frac{dy}{dt} = \frac{s_{1,a}(t) k_{1,a} - s_{3,a}(t) k_{3,a}}{y_3},
\end{equation}
\begin{equation}\nonumber
\frac{dz}{dt} = \frac{u'(t)}{2z(t)}, \quad u(t) = s_{1,a}(t)^2 - x(t)^2 - y(t)^2,
\end{equation}
\begin{equation}
u'(t) = 2 s_{1,a}(t) k_{1,a} - 2 x(t) \cdot \frac{dx}{dt} - 2 y(t) \cdot \frac{dy}{dt}.    
\end{equation}
The acceleration components, isolating the second derivatives, are:
\begin{equation}\nonumber
\frac{d^2 x}{dt^2} = \frac{k_{1,a}^2 - k_{2,a}^2}{x_2},
\end{equation}
\begin{equation}\nonumber
\frac{d^2 y}{dt^2} = \frac{k_{1,a}^2 - k_{3,a}^2}{y_3},
\end{equation}
\begin{equation}\label{eq:accel_z}
\frac{d^2 z}{dt^2} = \frac{u''(t) z(t)^2 - \frac{(u'(t))^2}{2}}{2z(t)^3},
\end{equation}
\begin{equation}
\begin{aligned}
u''(t) &= 2 k_{1,a}^2 - 2 \left( \left( \frac{dx}{dt} \right)^2 + x(t) \cdot \frac{d^2 x}{dt^2} \right) \\
&- 2 \left( \left( \frac{dy}{dt} \right)^2 + y(t) \cdot \frac{d^2 y}{dt^2} \right).
\end{aligned}
\end{equation}

The term $z(t)^{3}$ in the denominator of $\frac{d^2 z}{dt^2}$ reflects a potential $r^{-3}$ power law (Equations \ref{eq:r_vt} and \ref{eq:accel_z}). This power law is consistent with one of the bounds identified in Section \ref{sec:emergence_curved}.  To understand the scaling, we consider the two numerator terms, approximated under the conditions of Equations \ref{eq:equal_distances} and \ref{eq:equal_speeds}

\begin{equation}
\text{Term 1: }    u''(t)z(t)^2 \approx 2k_{1,a}^2z(t)^2
\end{equation}
\begin{equation}
\text{Term 2: } -\frac{(u'(t)^2)}{2} \approx -2k_{1,a}^2s_{1,a}(t)^2
\end{equation}

The first term is positive and proportional to the square of the vertical distance from the plane defined by $P_1$, $P_2$, $P_3$, while the second term is negative and proportional to the square of the distance from $P_1$. Since in our scenario $P_a$ is initially positioned above the center of mass with minimal in-plane motion, $s_{1.a}(t) > z(t)$ and the second term dominates, leading to negative acceleration consistent with the emergent metric's attractive dynamics (Section \ref{sec:emergence_curved}) and general relativity. The power law scaling depends on the relative contributions of these terms:
\begin{itemize}
    \item \textbf{Strong-field regime ($ |z(t)| \ll s_{1,a}(t) $)}: When $P_a$ is close to the x-y plane, the second term dominates, as $s_{1,a}(t)^2 >> z(t)^2$, yielding $\frac{d^2 z}{dt^2} \propto r^{-3}$, consistent with the strong-field limit in Section \ref{sec:emergence_curved}.
    \item \textbf{Weak-field regime ($|z(t)| \approx  s_{1,a}(t)$)}: When $P_a$ is further from the x-y plan, the vertical distance approaches the separation with $P_1$ and both terms contribute comparably, resulting in  $\frac{d^2 z}{dt^2} \propto r^{-2}$, aligning with the inverse-square law observed in gravitational-like effects and the weak-field limit. 
    \item \textbf{Other configurations}: An $r^{-1}$ scaling would require the first term to dominate ($z(t) >> s_{1,a}$), but this is precluded in configurations where $P_a$ is positioned above the center of mass, as $z(t) > s_{1,a}$ only occurs when $P_a$ is further from the center of mass than from $P_1$.
\end{itemize}

This analysis refines the power law bounds in Section \ref{sec:emergence_curved}, detailing the conditions for $r^{-3}$ and $r^{-2}$ scalings in this configuration, with $r^{-2}$ emerging as the dominant behavior at intermediate to large distances. Future research could explore alternative particle configurations or refine the mathematical solutions for sphere intersections to further validate these scalings (Section \ref{sec:discussion}) \cite{Cao2017,Franzmann2023}.

\section{Observer-dependent Metrics and Quantum non-Locality}\label{sec:observer_dependent}

In Quantum Spacetime Theory (QST), the emergent spacetime is constructed from \textbf{spacelike separations} $s_{i,j}(t)$, which are derived from the \textbf{mutual information} $I(S_i:S_j)(t)$ between quantum subsystems, as outlined in Section \ref{sec:QST} \cite{Cao2017, Cao2018, Franzmann2023}. A critical aspect of QST is its ability to model \textbf{quantum non-locality} where two particles exhibit high mutual information, corresponding to a small spacelike separation, even if their positions in a coordinate system defined by other particles suggest a large separation. This section provides a technical definition of how \textbf{entanglement} and its \textbf{non-local consequences} are modeled in QST, explores the observer-dependent nature of the emergent metric, and connects these concepts to established ideas such as the ER=EPR conjecture.

While spacelike separations are generally time-dependent, denoted $s_{i,j}(t)$ to reflect evolving mutual information (Sections \ref{sec:QST} and \ref{sec:emergence}), in this section, we analyze the geometric relationships at a fixed moment in time, using $s_{i,j}$ to denote the separation at a time t.

\subsection{Quantum entanglement and spacelike-separations}

Quantum entanglement is a phenomenon where the quantum states of two or more particles are correlated such that the state of one cannot be described independently of the others, regardless of their spatial separation \cite{Nielsen2000}. In QST, a higher degree of entanglement is modeled by higher mutual information $I(S_i:S_j)$ between two subsystems $S_i$ and $S_j$, corresponding to a small spacelike separation  $s_{i,j}(t) = l_{\text{type}} \Phi\left( I(S_i:S_j)(t)/I_{\text{max}} \right)$ (Equation \ref{eq:s_ia}) \cite{Cao2017, Cao2018, Franzmann2023}. 

This modeling is supported by the mathematics of quantum information theory, where mutual information is defined as:

\begin{equation}\nonumber
I(S_i:S_j) = S_i + S_j - S_{i \cup j},
\end{equation}

where $S_i$, $S_j$, and $S_{i \cup j}$ are the von Neumann entropies of subsystems $S_i$, $S_j$, and their union, respectively, evaluated at a given time t \cite{Nielsen2000}. For entangled particles, $I(S_i:S_j)$ is large, reducing $s_{i,j}$, which aligns with the emergent geometry of QST. This framework is consistent with holographic theories, where entanglement entropy is related to geometric quantities via the Ryu-Takayanagi formula:

\begin{equation}\nonumber
S_A = \frac{\text{Area}(\gamma_A)}{4 G_N},
\end{equation}

where $S_A$ is the entanglement entropy of subsystem $A$, $\gamma_A$ is a minimal surface in the bulk spacetime, and $G_N$ is Newton's gravitational constant \cite{Ryu2006}. In QST, the inverse relationship $s_{i,j} \propto 1 / I(S_i:S_j)$ directly ties quantum correlations to spatial distances, providing a geometric interpretation of entanglement.

\subsection{Non-Locality and the ER=EPR Conjecture}

To illustrate modeling quantum non-locality we consider three particles $P_{1}, P_{2}$ and $P_{3}$ where $s_{1,3} = 10$ and $s_{1,2} = s_{2,3} = 1$. In a flat coordinate space where $s_{1,3}$ and $s_{1,2}$ are accurately represented, the coordinate distance between $P_{2}$ and $P_{3}$ would be $d(P_2,P_3) \geq s_{1,3} - s_{1,2} = 9$ (see Fig \ref{fig:entanglement}a). However, the small $s_{2,3}$, which arises from high $I(S_2:S_3)$, indicates that $P_{2}$ and $P_{3}$ are highly entangled and thus "close" in the emergent spacetime, despite their coordinate positions suggesting a larger separation. This duality—--large coordinate separation versus small metric distance—--captures the non-local nature of entanglement, where quantum correlations transcend classical spatial constraints.

\begin{figure}[ht]
\centering
\begin{tikzpicture}

  \node at (2.0,6.1) {(a) $d(P_2,P_3) > s_{2,3}$.};
  
  \fill[black] (0,5) circle (0.1cm) node[above] {$P_1$};
  \fill[black] (4,5) circle (0.1cm) node[above] {$P_3$};
  \fill[black] (0.4,5) circle (0.1cm) node[above] {$P_2$};
  
  \draw[-, black, thick] (0,5) -- (4.0,5) node[midway,below]{$s_{1,3} = 10$};
  \draw[-, black, thick] (0,5) -- (0.4,5) node[below, yshift=-0.1cm]{$s_{1,2} = 1$};
  \draw[-, black, dashed, thick] (0.4,5)  .. controls (1.4,5.5) and (3,5.5) .. (4,5) node[midway,above]{$s_{2,3}=1$};

  \node at (2.0,3.6) {(b) $d(P_1,P_2) > s_{1,2}$.};
  
  \fill[black] (0,2.5) circle (0.1cm) node[above] {$P_1$};
  \fill[black] (4,2.5) circle (0.1cm) node[above] {$P_3$};
  \fill[black] (3.6,2.5) circle (0.1cm) node[above] {$P_2$};
  
  \draw[-, black,thick] (0,2.5) -- (4.0,2.5) node[midway,below]{$s_{1,3} = 10$};
  
  \draw[-, black, dashed,thick] (0,2.5)  .. controls (1,3) and (3,3) .. (3.6,2.5) node[midway, above]{$s_{1,2} = 1$};
  \draw[-, black,thick] (4.0,2.5) -- (3.6,2.5) node[below, xshift=0.5cm]{$s_{2,3}=1$};

\node at (2,1.3) {(c) $d(P_1,P_3)  < s_{1,3}$.};
  
  \fill[black] (1.6,-0.4) circle (0.1cm) node[left] {$P_1$};
  \fill[black] (2.4,-0.4) circle (0.1cm) node[right] {$P_3$};
  \fill[black] (2.0,-0.4) circle (0.1cm) node[above] {$P_2$};
  
  \draw[-, black,thick] (1.6,-0.4) -- (2.0,-0.4);
  \draw[-, black,thick] (2.0,-0.4) -- (2.4,-0.4) node[midway,below, yshift=-0.1cm]{$s_{1,2} = s_{2,3}=1$};

  \draw[black, -, dashed,thick] (1.6,-0.4)  .. controls (-0.4,1.3) and (4.4,1.3) .. (2.4,-0.4) node[midway, below,yshift=-0.2cm] {$s_{1,3} = 10$};

\end{tikzpicture}
\caption{It is not possible for a flat metric to accurately represent three particles $P_i$ where their spacelike intervals do not satisfy the triangle inequality (e.g. $s_{i,j} > s_{i,k} + s_{k,j}$), such as $s_{1,3} = 10$ and $s_{1,2} = s_{2,3} = 1$, since $10 > 1 + 1$. As shown, the coordinate distance $d()$ for one of the particle pairs $P_i,P_j$ must be more (a or b) or less (c) than the corresponding interval $s_{i,j}$.}

\label{fig:entanglement}
\end{figure}

This scenario, where the coordinate distance $d(P_i,P_j) > s_{i,j}$ (Fig. \ref{fig:entanglement}a and b), finds a strong parallel in the ER=EPR conjecture, proposed by Maldacena and Susskind \cite{Maldacena2013, Susskind2016}. This conjecture posits that entangled particles are connected by Einstein-Rosen (ER) bridges, or wormholes, which act as shortcuts in spacetime, effectively reducing the distance between the particles. For example, in our scenario where $d(P_2,P3) \geq 9$ but $s_{2,3}  = 1$ (Fig. \ref{fig:entanglement}a), the small $s_{2,3}$ suggests that $P_2$ and $P_3$ are connected by a structure analogous to a wormhole, allowing their entangled states to influence each other more strongly than their coordinate separation would imply. This is consistent with the ER=EPR idea, where entanglement creates a non-trivial spacetime geometry, and supports QST's approach to modeling entanglement as a fundamental driver of spacetime structure.

Research by van Raamsdonk, 2010 further reinforces this connection, arguing that spacetime connectivity emerges from quantum entanglement, with unentangled systems corresponding to disconnected spacetime regions \cite{VanRaamsdonk2010}. In QST, the high $I(S_2:S_3)$ ensures that $P_2$ and $P_3$ remain connected in the emergent spacetime, mirroring these holographic principles.

\subsection{Observer-Dependence Metrics}

For particles $P_1$, $P_2$, and $P_3$, their spacelike separations $s_{i,j}$ ($s_{1,3} = 10, s_{1,2} = s_{2,3}=3$) violate the triangle inequality, requiring one of the coordinate distances $d(P_i,P_j)$ to either exceed $s_{i,j}$ (Figure \ref{fig:entanglement}a, b), supporting quantum non-locality via geometric shortcuts akin to ER bridges \cite{Maldacena2013}, or be less than $s_{i,j}$ (Figure \ref{fig:entanglement}c). The metric graph's shortest-path distance function enforces $d(P_i,P_j) \leq s_{i,j}$, satisfying the triangle inequality but precluding non-locality \cite{Cao2017,Franzmann2023}. To allow for non-locality, we propose observer-dependent metrics $g_{\mu\nu}^{(k)}(t)$,  defined by a reference subsystem $S_k$, analogous to relational quantum mechanics \cite{Rovelli1996}. Coordinate distances $d_k(P_i,P_j)$ in the observer's metric $g_{\mu\nu}^{(k)}(t)$ are determined by conditional mutual information $I(S_i:S_j|S_k)$, which reflects $S_k$'s entanglement context, while the spacelike separations $s_{i,j}$ fixed by $I(S_i:S_j)$, remain invariant across observers. For example, observer $S_1$ prioritizes $s_{1,2}$, $s_{1,3}$, yielding $d_1(P_2,P_3) > s_{2,3}$, while $S_2$ prioritizes $s_{2,3}$. These metrics may induce entanglement correlations, potentially testable through non-locality experiments or CMB observations, though requiring refinement (Section \ref{sec:discussion})

\section{Discussion}\label{sec:discussion}

Quantum Spacetime Theory (QST) advances emergent spacetime by deriving geometric properties from quantum subsystem relationships, quantified by mutual information $I(S_i:S_j)$ (Section \ref{sec:QST}). Unlike general relativity (GR), where curvature arises from mass-energy via Einstein’s field equations, QST derives spacetime curvature from relational spacelike separations $s_{i,j}(t)$, demonstrating gravitational-like acceleration effects ranging from $r^{-1}$ to $r^{-3}$, including the inverse-square law (Section \ref{sec:emergence_curved}). Building on the metric graph framework \cite{Cao2017, Franzmann2023}, QST replaces shortest-path distances with sphere intersections and introduces observer-dependent emergent metrics to satisfy the triangle inequality, accommodating quantum non-locality consistent with the ER=EPR conjecture (Section \ref{sec:observer_dependent}) \cite{Maldacena2013}.

QST’s core principle is that spacelike separations are derived from quantum subsystem correlations, not tied to a specific mapping. We propose:
\begin{equation}\nonumber
s_{i,j}(t) = l_{\text{type}} \Phi\left( \frac{I(S_i:S_j)(t)}{I_{\text{max}}} \right),
\end{equation}
where $l_{\text{type}}$ is a type-specific length scale (e.g., $10^{-9} \, \text{m}$ for qubits, $l_P \approx 1.616 \times 10^{-35} \, \text{m}$ for fields) and $\Phi(x)$ (e.g., $-\log(x)$) is monotonically decreasing \cite{Ryu2006}. This ensures $s_{i,j} = 0$ for fully entangled subsystems ($ I(S_i:S_j) = I_{\text{max}} $), effectively one quantum entity, and increases as subsystems become more distinct (lower $I(S_i:S_j)$) \cite{Nielsen2000}. However, QST’s validity rests on the general relationship, not this specific form. Alternative mappings, such as $s_{i,j} = \kappa \Phi(S_i, S_j)$, should be explored if they maintain these characteristics.

Future research should test the proposed equation, validating $\Phi(x)$ and $l_{\text{type}}$, and investigate alternatives across systems, from qubit experiments to cosmological contexts like cosmic web formation \cite{Planck2018, Springel2005}. Theoretical refinements could align QST with holographic principles or loop quantum gravity \cite{Ryu2006, Ashtekar2004}. Extending sphere intersections to dynamic $s_{i,j}(t)$ and numerical solutions to address computational complexity, will enhance cosmological applications. Refining observer-dependent metrics using conditional mutual information (Section \ref{sec:observer_dependent}) will enhance non-locality modeling, potentially observable in non-locality tests or CMB data \cite{Bell1964, Planck2018}. QST’s presentist framework, deriving curved trajectories from relational dynamics (Fig. \ref{fig:qst_vs_gr}), complements GR, offering a quantum-geometric perspective to unify quantum mechanics and gravity.

\begin{figure}[ht]
\centering
\begin{tikzpicture}
  \node at (0.5,5.4) {(a) GR};
  \foreach \x in {-0.5,0,...,1.5} {
    \draw[gray, dashed] (\x,3) .. controls (\x+0.2,4) and (\x-0.2,4) .. (\x,5);
  }
  \foreach \y in {3,3.5,...,5} {
    \draw[gray, dashed] (-0.5,\y) .. controls (0.5,{\y+0.2}) and (0.5,{\y-0.2}) .. (1.5,\y);
  }
  \fill[black] (0.5,4.16) circle (0.1cm) node[below] {$T_{\mu\nu}$};
  \draw[black, thick, ->] (-0.5,3.5) .. controls (0.3,4.5) and (1.5,4.5) .. (1.5,3.5) node[midway, above] {Geodesic};
  \node at (0.5,3.2) {$G_{\mu\nu}$};

  \node at (0.5,2.4) {(b) QST};
  \fill[black] (-0.6,0) circle (0.1cm) node[below] {$P_{a}$};
  \fill[black] (2,0) circle (0.1cm) node[below] {$P_2$};
  \fill[black] (1,1.73) circle (0.1cm) node[above] {$P_1$};
  \draw[-] (-0.6,0) -- (2,0) node[midway, below] {$s_{a,2}(t)$};
  \draw[-] (-0.6,0) -- (1,1.73) node[midway, left] {$s_{a,1}(t)$};
  \draw[-] (2,0) -- (1,1.73) node[midway, right,yshift=0.3cm,xshift=-0.2cm] {$s_{1,2}(t)$};
  \draw[red, dashed, thick, ->] (-0.6,0) .. controls (0.4,0.4) and (0.8,0.8) .. (1.2,1.4) node[midway,right] {$P_a$ trajectory};
    \draw[-, red] (0.2,0.4) -- (1,1.73);
    \draw[-, red] (0.2,0.4) -- (2,0);
\end{tikzpicture}
\caption{Curvature in GR and novel QST. (a) GR’s spacetime curvature from Einstein’s field equations, $G_{\mu\nu} + \Lambda g_{\mu\nu} = 8\pi T_{\mu\nu}$, driven by mass-energy. (b) QST’s observer-independent spacelike separations $s_{i,j}(t)$ (straight edges) and curved trajectory of $P_a$ due to relational dynamics (Section \ref{sec:emergence_curved}).}
\label{fig:qst_vs_gr}
\end{figure}

\section{Conclusion}

The proposed QST reinterprets spacetime as an emergent structure, advancing quantum gravity through a relational framework driven by quantum constraints. We reviewed the metric graph framework, where a 3D metric space emerges from quantum entanglement, with edge weights defined by mutual information \cite{Cao2017, Cao2018, Franzmann2023}. Building on this, we derived a gravitational-like acceleration effect for a particle $P_a$, yielding a range from $r^{-1}$ to $r^{-3}$ with $r^{-2}$ (inverse-square law) within this range, driven by mutual information constraints rather than classical mass. The novel QST introduces observer-dependent spacetime metrics, positioning particles via sphere intersections, with non-intersecting cases leading to geometric shortcuts, consistent with quantum non-locality and the ER=EPR conjecture, and multiple metrics. QST shares features with general relativity, including curved spacetime, but posits that spacetime metrics are a direct consequence of observer-independent spacelike separations $s_{i,j}(t)$, with relational laws $\mathcal{L}(\{ s_{i,j}(t) \}, \{ \dot{s}_{i,j}(t) \}, \ldots)$. QST also posits a presentist perspective, contrasting GR’s eternalist 4D metric.

The direct emergence of the spacetime metric from spacelike separations, without intermediaries like distance functions, distinguishes QST from prior frameworks \cite{Cao2017, Cao2018, Franzmann2023}. The inevitability of curved space due to relational constraints offers a novel framework for unifying quantum mechanics and general relativity. However, the quantum-to-geometric correspondence remains incomplete \cite{Franzmann2023}. Future research should refine this mapping, explore experimental tests of QST’s predictions, and investigate implications for cosmic evolution, potentially advancing our understanding of spacetime’s quantum origins.

\section*{Declarations}
The author declares no competing financial or non-financial interests.

\bibliography{references}

\begin{thebibliography}{27}%
\makeatletter
\providecommand \@ifxundefined [1]{%
 \@ifx{#1\undefined}
}%
\providecommand \@ifnum [1]{%
 \ifnum #1\expandafter \@firstoftwo
 \else \expandafter \@secondoftwo
 \fi
}%
\providecommand \@ifx [1]{%
 \ifx #1\expandafter \@firstoftwo
 \else \expandafter \@secondoftwo
 \fi
}%
\providecommand \natexlab [1]{#1}%
\providecommand \enquote  [1]{``#1''}%
\providecommand \bibnamefont  [1]{#1}%
\providecommand \bibfnamefont [1]{#1}%
\providecommand \citenamefont [1]{#1}%
\providecommand \href@noop [0]{\@secondoftwo}%
\providecommand \href [0]{\begingroup \@sanitize@url \@href}%
\providecommand \@href[1]{\@@startlink{#1}\@@href}%
\providecommand \@@href[1]{\endgroup#1\@@endlink}%
\providecommand \@sanitize@url [0]{\catcode `\\12\catcode `\$12\catcode `\&12\catcode `\#12\catcode `\^12\catcode `\_12\catcode `\%12\relax}%
\providecommand \@@startlink[1]{}%
\providecommand \@@endlink[0]{}%
\providecommand \url  [0]{\begingroup\@sanitize@url \@url }%
\providecommand \@url [1]{\endgroup\@href {#1}{\urlprefix }}%
\providecommand \urlprefix  [0]{URL }%
\providecommand \Eprint [0]{\href }%
\providecommand \doibase [0]{https://doi.org/}%
\providecommand \selectlanguage [0]{\@gobble}%
\providecommand \bibinfo  [0]{\@secondoftwo}%
\providecommand \bibfield  [0]{\@secondoftwo}%
\providecommand \translation [1]{[#1]}%
\providecommand \BibitemOpen [0]{}%
\providecommand \bibitemStop [0]{}%
\providecommand \bibitemNoStop [0]{.\EOS\space}%
\providecommand \EOS [0]{\spacefactor3000\relax}%
\providecommand \BibitemShut  [1]{\csname bibitem#1\endcsname}%
\let\auto@bib@innerbib\@empty
\bibitem [{\citenamefont {Ashtekar}\ and\ \citenamefont {Lewandowski}(2004)}]{Ashtekar2004}%
  \BibitemOpen
  \bibfield  {author} {\bibinfo {author} {\bibfnamefont {A.}~\bibnamefont {Ashtekar}}\ and\ \bibinfo {author} {\bibfnamefont {J.}~\bibnamefont {Lewandowski}},\ }\bibfield  {title} {\bibinfo {title} {Background independent quantum gravity: a status report},\ }\href {https://doi.org/10.1088/0264-9381/21/15/R01} {\bibfield  {journal} {\bibinfo  {journal} {Classical and Quantum Gravity}\ }\textbf {\bibinfo {volume} {21}},\ \bibinfo {pages} {R53} (\bibinfo {year} {2004})}\BibitemShut {NoStop}%
\bibitem [{\citenamefont {Maldacena}(1998)}]{Maldacena1998}%
  \BibitemOpen
  \bibfield  {author} {\bibinfo {author} {\bibfnamefont {J.}~\bibnamefont {Maldacena}},\ }\bibfield  {title} {\bibinfo {title} {The large $n$ limit of superconformal field theories and supergravity},\ }\href {https://doi.org/10.4310/ATMP.1998.V2.N2.A1} {\bibfield  {journal} {\bibinfo  {journal} {Advances in Theoretical and Mathematical Physics}\ }\textbf {\bibinfo {volume} {2}},\ \bibinfo {pages} {231} (\bibinfo {year} {1998})}\BibitemShut {NoStop}%
\bibitem [{\citenamefont {Ryu}\ and\ \citenamefont {Takayanagi}(2006)}]{Ryu2006}%
  \BibitemOpen
  \bibfield  {author} {\bibinfo {author} {\bibfnamefont {S.}~\bibnamefont {Ryu}}\ and\ \bibinfo {author} {\bibfnamefont {T.}~\bibnamefont {Takayanagi}},\ }\bibfield  {title} {\bibinfo {title} {Holographic derivation of entanglement entropy from the anti-de sitter space/conformal field theory correspondence},\ }\href {https://doi.org/10.1103/PHYSREVLETT.96.181602} {\bibfield  {journal} {\bibinfo  {journal} {Physical Review Letters}\ }\textbf {\bibinfo {volume} {96}},\ \bibinfo {pages} {181602} (\bibinfo {year} {2006})}\BibitemShut {NoStop}%
\bibitem [{\citenamefont {van Raamsdonk}(2010)}]{VanRaamsdonk2010}%
  \BibitemOpen
  \bibfield  {author} {\bibinfo {author} {\bibfnamefont {M.}~\bibnamefont {van Raamsdonk}},\ }\bibfield  {title} {\bibinfo {title} {Building up spacetime with quantum entanglement},\ }\href {https://doi.org/10.1007/S10714-010-1034-0} {\bibfield  {journal} {\bibinfo  {journal} {General Relativity and Gravitation}\ }\textbf {\bibinfo {volume} {42}},\ \bibinfo {pages} {2323} (\bibinfo {year} {2010})}\BibitemShut {NoStop}%
\bibitem [{\citenamefont {Rovelli}(1996)}]{Rovelli1996}%
  \BibitemOpen
  \bibfield  {author} {\bibinfo {author} {\bibfnamefont {C.}~\bibnamefont {Rovelli}},\ }\bibfield  {title} {\bibinfo {title} {Relational quantum mechanics},\ }\href {https://doi.org/10.1007/BF02302261} {\bibfield  {journal} {\bibinfo  {journal} {International Journal of Theoretical Physics}\ }\textbf {\bibinfo {volume} {35}},\ \bibinfo {pages} {1637} (\bibinfo {year} {1996})}\BibitemShut {NoStop}%
\bibitem [{\citenamefont {Swingle}(2012)}]{Swingle2012}%
  \BibitemOpen
  \bibfield  {author} {\bibinfo {author} {\bibfnamefont {B.}~\bibnamefont {Swingle}},\ }\bibfield  {title} {\bibinfo {title} {Entanglement renormalization and holography},\ }\href {https://doi.org/10.1103/PHYSREVD.86.065007} {\bibfield  {journal} {\bibinfo  {journal} {Physical Review D}\ }\textbf {\bibinfo {volume} {86}},\ \bibinfo {pages} {065007} (\bibinfo {year} {2012})}\BibitemShut {NoStop}%
\bibitem [{Note1()}]{Note1}%
  \BibitemOpen
  \bibinfo {note} {See also \cite {Pastawski2015, Qi2018, Boldis2024} for related entanglement-geometry studies.}\BibitemShut {Stop}%
\bibitem [{\citenamefont {Cao}\ \emph {et~al.}(2017)\citenamefont {Cao}, \citenamefont {Carroll},\ and\ \citenamefont {Michalakis}}]{Cao2017}%
  \BibitemOpen
  \bibfield  {author} {\bibinfo {author} {\bibfnamefont {C.}~\bibnamefont {Cao}}, \bibinfo {author} {\bibfnamefont {S.~M.}\ \bibnamefont {Carroll}},\ and\ \bibinfo {author} {\bibfnamefont {S.}~\bibnamefont {Michalakis}},\ }\bibfield  {title} {\bibinfo {title} {Space from hilbert space: Recovering geometry from bulk entanglement},\ }\href {https://doi.org/10.1103/PHYSREVD.95.024031} {\bibfield  {journal} {\bibinfo  {journal} {Physical Review D}\ }\textbf {\bibinfo {volume} {95}},\ \bibinfo {pages} {024031} (\bibinfo {year} {2017})}\BibitemShut {NoStop}%
\bibitem [{\citenamefont {Cao}\ and\ \citenamefont {Carroll}(2018)}]{Cao2018}%
  \BibitemOpen
  \bibfield  {author} {\bibinfo {author} {\bibfnamefont {C.}~\bibnamefont {Cao}}\ and\ \bibinfo {author} {\bibfnamefont {S.~M.}\ \bibnamefont {Carroll}},\ }\bibfield  {title} {\bibinfo {title} {Bulk entanglement gravity without a boundary: Towards finding einstein's equation in hilbert space},\ }\href {https://doi.org/10.1103/PhysRevD.97.086003} {\bibfield  {journal} {\bibinfo  {journal} {Physical Review D}\ }\textbf {\bibinfo {volume} {97}},\ \bibinfo {pages} {086003} (\bibinfo {year} {2018})}\BibitemShut {NoStop}%
\bibitem [{\citenamefont {Franzmann}\ \emph {et~al.}(2023)\citenamefont {Franzmann}, \citenamefont {Jovancic},\ and\ \citenamefont {Lawson}}]{Franzmann2023}%
  \BibitemOpen
  \bibfield  {author} {\bibinfo {author} {\bibfnamefont {G.}~\bibnamefont {Franzmann}}, \bibinfo {author} {\bibfnamefont {S.~M.~D.}\ \bibnamefont {Jovancic}},\ and\ \bibinfo {author} {\bibfnamefont {M.}~\bibnamefont {Lawson}},\ }\bibfield  {title} {\bibinfo {title} {Relative distance of entangled systems in emergent spacetime scenarios},\ }\href {https://doi.org/10.1103/PhysRevD.107.066008} {\bibfield  {journal} {\bibinfo  {journal} {Physical Review D}\ }\textbf {\bibinfo {volume} {107}},\ \bibinfo {pages} {066008} (\bibinfo {year} {2023})}\BibitemShut {NoStop}%
\bibitem [{\citenamefont {Maldacena}\ and\ \citenamefont {Susskind}(2013)}]{Maldacena2013}%
  \BibitemOpen
  \bibfield  {author} {\bibinfo {author} {\bibfnamefont {J.}~\bibnamefont {Maldacena}}\ and\ \bibinfo {author} {\bibfnamefont {L.}~\bibnamefont {Susskind}},\ }\bibfield  {title} {\bibinfo {title} {Cool horizons for entangled black holes},\ }\href {https://doi.org/10.1002/PROP.201300020} {\bibfield  {journal} {\bibinfo  {journal} {Fortschritte der Physik}\ }\textbf {\bibinfo {volume} {61}},\ \bibinfo {pages} {781} (\bibinfo {year} {2013})}\BibitemShut {NoStop}%
\bibitem [{\citenamefont {Nielsen}\ and\ \citenamefont {Chuang}(2000)}]{Nielsen2000}%
  \BibitemOpen
  \bibfield  {author} {\bibinfo {author} {\bibfnamefont {M.~A.}\ \bibnamefont {Nielsen}}\ and\ \bibinfo {author} {\bibfnamefont {I.~L.}\ \bibnamefont {Chuang}},\ }\href@noop {} {\emph {\bibinfo {title} {Quantum Computation and Quantum Information}}}\ (\bibinfo  {publisher} {Cambridge University Press},\ \bibinfo {address} {Cambridge, England},\ \bibinfo {year} {2000})\BibitemShut {NoStop}%
\bibitem [{\citenamefont {Carroll}(2004)}]{Carroll2004spacetime}%
  \BibitemOpen
  \bibfield  {author} {\bibinfo {author} {\bibfnamefont {S.~M.}\ \bibnamefont {Carroll}},\ }\href@noop {} {\emph {\bibinfo {title} {Spacetime and Geometry: An Introduction to General Relativity}}}\ (\bibinfo  {publisher} {Addison-Wesley},\ \bibinfo {address} {San Francisco},\ \bibinfo {year} {2004})\BibitemShut {NoStop}%
\bibitem [{\citenamefont {Srednicki}(1993)}]{Srednicki1993}%
  \BibitemOpen
  \bibfield  {author} {\bibinfo {author} {\bibfnamefont {M.}~\bibnamefont {Srednicki}},\ }\bibfield  {title} {\bibinfo {title} {Entropy and area},\ }\href {https://doi.org/10.1103/PhysRevLett.71.666} {\bibfield  {journal} {\bibinfo  {journal} {Physical Review Letters}\ }\textbf {\bibinfo {volume} {71}},\ \bibinfo {pages} {666} (\bibinfo {year} {1993})}\BibitemShut {NoStop}%
\bibitem [{\citenamefont {Pastawski}\ \emph {et~al.}(2015)\citenamefont {Pastawski}, \citenamefont {Yoshida}, \citenamefont {Harlow},\ and\ \citenamefont {Preskill}}]{Pastawski2015}%
  \BibitemOpen
  \bibfield  {author} {\bibinfo {author} {\bibfnamefont {F.}~\bibnamefont {Pastawski}}, \bibinfo {author} {\bibfnamefont {B.}~\bibnamefont {Yoshida}}, \bibinfo {author} {\bibfnamefont {D.}~\bibnamefont {Harlow}},\ and\ \bibinfo {author} {\bibfnamefont {J.}~\bibnamefont {Preskill}},\ }\bibfield  {title} {\bibinfo {title} {Holographic quantum error-correcting codes: toy models for the bulk/boundary correspondence},\ }\href {https://doi.org/10.1007/JHEP06(2015)149} {\bibfield  {journal} {\bibinfo  {journal} {Journal of High Energy Physics}\ }\textbf {\bibinfo {volume} {2015}},\ \bibinfo {pages} {1} (\bibinfo {year} {2015})}\BibitemShut {NoStop}%
\bibitem [{\citenamefont {Qi}(2018)}]{Qi2018}%
  \BibitemOpen
  \bibfield  {author} {\bibinfo {author} {\bibfnamefont {X.~L.}\ \bibnamefont {Qi}},\ }\bibfield  {title} {\bibinfo {title} {Does gravity come from quantum information?},\ }\href {https://doi.org/10.1038/s41567-018-0297-3} {\bibfield  {journal} {\bibinfo  {journal} {Nature Physics}\ }\textbf {\bibinfo {volume} {14}},\ \bibinfo {pages} {984} (\bibinfo {year} {2018})}\BibitemShut {NoStop}%
\bibitem [{\citenamefont {{Planck Collaboration}}(2020)}]{Planck2018}%
  \BibitemOpen
  \bibfield  {author} {\bibinfo {author} {\bibnamefont {{Planck Collaboration}}},\ }\bibfield  {title} {\bibinfo {title} {Planck 2018 results. vi. cosmological parameters},\ }\href {https://doi.org/10.1051/0004-6361/201833910} {\bibfield  {journal} {\bibinfo  {journal} {Astronomy \& Astrophysics}\ }\textbf {\bibinfo {volume} {641}},\ \bibinfo {pages} {A6} (\bibinfo {year} {2020})}\BibitemShut {NoStop}%
\bibitem [{\citenamefont {Jacobson}(1995)}]{Jacobson1995}%
  \BibitemOpen
  \bibfield  {author} {\bibinfo {author} {\bibfnamefont {T.}~\bibnamefont {Jacobson}},\ }\bibfield  {title} {\bibinfo {title} {Thermodynamics of spacetime: The einstein equation of state},\ }\href {https://doi.org/10.1103/PhysRevLett.75.1260} {\bibfield  {journal} {\bibinfo  {journal} {Physical Review Letters}\ }\textbf {\bibinfo {volume} {75}},\ \bibinfo {pages} {1260} (\bibinfo {year} {1995})}\BibitemShut {NoStop}%
\bibitem [{\citenamefont {Goldstein}\ \emph {et~al.}(2002)\citenamefont {Goldstein}, \citenamefont {Poole},\ and\ \citenamefont {Safko}}]{goldstein2002classical}%
  \BibitemOpen
  \bibfield  {author} {\bibinfo {author} {\bibfnamefont {H.}~\bibnamefont {Goldstein}}, \bibinfo {author} {\bibfnamefont {C.}~\bibnamefont {Poole}},\ and\ \bibinfo {author} {\bibfnamefont {J.}~\bibnamefont {Safko}},\ }\href@noop {} {\emph {\bibinfo {title} {Classical Mechanics}}},\ \bibinfo {edition} {3rd}\ ed.\ (\bibinfo  {publisher} {Addison-Wesley},\ \bibinfo {address} {San Francisco},\ \bibinfo {year} {2002})\BibitemShut {NoStop}%
\bibitem [{\citenamefont {Misner}\ \emph {et~al.}(1973)\citenamefont {Misner}, \citenamefont {Thorne},\ and\ \citenamefont {Wheeler}}]{misner1973gravitation}%
  \BibitemOpen
  \bibfield  {author} {\bibinfo {author} {\bibfnamefont {C.~W.}\ \bibnamefont {Misner}}, \bibinfo {author} {\bibfnamefont {K.~S.}\ \bibnamefont {Thorne}},\ and\ \bibinfo {author} {\bibfnamefont {J.~A.}\ \bibnamefont {Wheeler}},\ }\href@noop {} {\emph {\bibinfo {title} {Gravitation}}}\ (\bibinfo  {publisher} {W. H. Freeman},\ \bibinfo {address} {San Francisco},\ \bibinfo {year} {1973})\BibitemShut {NoStop}%
\bibitem [{\citenamefont {Rahman}(2012)}]{rahman2012beyond}%
  \BibitemOpen
  \bibfield  {author} {\bibinfo {author} {\bibfnamefont {M.~Z.}\ \bibnamefont {Rahman}},\ }\bibfield  {title} {\bibinfo {title} {Beyond trilateration: Gps positioning geometry and analytical accuracy},\ }in\ \href {https://doi.org/10.5772/30602} {\emph {\bibinfo {booktitle} {Global Navigation Satellite Systems - From Stellar to Terrestrial}}},\ \bibinfo {editor} {edited by\ \bibinfo {editor} {\bibfnamefont {A.}~\bibnamefont {V{\"a}l{\"a}j{\"a}rvi}}}\ (\bibinfo  {publisher} {InTech},\ \bibinfo {address} {London, England},\ \bibinfo {year} {2012})\BibitemShut {NoStop}%
\bibitem [{\citenamefont {Abdullah}\ \emph {et~al.}(2020)\citenamefont {Abdullah}, \citenamefont {Jusoh}, \citenamefont {Ismail}, \citenamefont {Ali},\ and\ \citenamefont {Jusoh}}]{abdullah2020position}%
  \BibitemOpen
  \bibfield  {author} {\bibinfo {author} {\bibfnamefont {N.~A.}\ \bibnamefont {Abdullah}}, \bibinfo {author} {\bibfnamefont {M.~A.}\ \bibnamefont {Jusoh}}, \bibinfo {author} {\bibfnamefont {A.}~\bibnamefont {Ismail}}, \bibinfo {author} {\bibfnamefont {M.}~\bibnamefont {Ali}},\ and\ \bibinfo {author} {\bibfnamefont {N.~A.}\ \bibnamefont {Jusoh}},\ }\bibfield  {title} {\bibinfo {title} {Position estimation using trilateration based on toa/rss and aoa measurement},\ }\href {https://doi.org/10.1088/1742-6596/1773/1/012002} {\bibfield  {journal} {\bibinfo  {journal} {Journal of Physics: Conference Series}\ }\textbf {\bibinfo {volume} {1529}},\ \bibinfo {pages} {012012} (\bibinfo {year} {2020})}\BibitemShut {NoStop}%
\bibitem [{\citenamefont {Li}\ and\ \citenamefont {Bowyer}(1997)}]{li1997torus}%
  \BibitemOpen
  \bibfield  {author} {\bibinfo {author} {\bibfnamefont {Q.}~\bibnamefont {Li}}\ and\ \bibinfo {author} {\bibfnamefont {A.}~\bibnamefont {Bowyer}},\ }\bibfield  {title} {\bibinfo {title} {Torus/sphere intersection based on a configuration space approach},\ }\href {https://doi.org/10.1006/GMIP.1997.0451} {\bibfield  {journal} {\bibinfo  {journal} {Computer-Aided Design}\ }\textbf {\bibinfo {volume} {29}},\ \bibinfo {pages} {847} (\bibinfo {year} {1997})}\BibitemShut {NoStop}%
\bibitem [{\citenamefont {Susskind}(2016)}]{Susskind2016}%
  \BibitemOpen
  \bibfield  {author} {\bibinfo {author} {\bibfnamefont {L.}~\bibnamefont {Susskind}},\ }\bibfield  {title} {\bibinfo {title} {Entanglement is not enough},\ }\href {https://doi.org/10.1002/prop.201500095} {\bibfield  {journal} {\bibinfo  {journal} {Fortschritte der Physik}\ }\textbf {\bibinfo {volume} {64}},\ \bibinfo {pages} {49} (\bibinfo {year} {2016})}\BibitemShut {NoStop}%
\bibitem [{\citenamefont {Springel}\ \emph {et~al.}(2005)\citenamefont {Springel}, \citenamefont {White}, \citenamefont {Jenkins}, \citenamefont {Frenk}, \citenamefont {Yoshida}, \citenamefont {Gao}, \citenamefont {Navarro}, \citenamefont {Thacker}, \citenamefont {Croton}, \citenamefont {Helly}, \citenamefont {Peacock}, \citenamefont {Cole}, \citenamefont {Thomas}, \citenamefont {Couchman}, \citenamefont {Evrard}, \citenamefont {Colberg},\ and\ \citenamefont {Pearce}}]{Springel2005}%
  \BibitemOpen
  \bibfield  {author} {\bibinfo {author} {\bibfnamefont {V.}~\bibnamefont {Springel}}, \bibinfo {author} {\bibfnamefont {S.~D.~M.}\ \bibnamefont {White}}, \bibinfo {author} {\bibfnamefont {A.}~\bibnamefont {Jenkins}}, \bibinfo {author} {\bibfnamefont {C.~S.}\ \bibnamefont {Frenk}}, \bibinfo {author} {\bibfnamefont {N.}~\bibnamefont {Yoshida}}, \bibinfo {author} {\bibfnamefont {L.}~\bibnamefont {Gao}}, \bibinfo {author} {\bibfnamefont {J.}~\bibnamefont {Navarro}}, \bibinfo {author} {\bibfnamefont {R.}~\bibnamefont {Thacker}}, \bibinfo {author} {\bibfnamefont {D.}~\bibnamefont {Croton}}, \bibinfo {author} {\bibfnamefont {J.}~\bibnamefont {Helly}}, \bibinfo {author} {\bibfnamefont {J.~A.}\ \bibnamefont {Peacock}}, \bibinfo {author} {\bibfnamefont {S.}~\bibnamefont {Cole}}, \bibinfo {author} {\bibfnamefont {P.}~\bibnamefont {Thomas}}, \bibinfo {author} {\bibfnamefont {H.}~\bibnamefont {Couchman}}, \bibinfo {author} {\bibfnamefont {A.}~\bibnamefont {Evrard}}, \bibinfo {author} {\bibfnamefont {J.}~\bibnamefont
  {Colberg}},\ and\ \bibinfo {author} {\bibfnamefont {F.}~\bibnamefont {Pearce}},\ }\bibfield  {title} {\bibinfo {title} {Simulations of the formation, evolution and clustering of galaxies and quasars},\ }\href {https://doi.org/10.1038/nature03597} {\bibfield  {journal} {\bibinfo  {journal} {Nature}\ }\textbf {\bibinfo {volume} {435}},\ \bibinfo {pages} {629} (\bibinfo {year} {2005})}\BibitemShut {NoStop}%
\bibitem [{\citenamefont {Bell}(1964)}]{Bell1964}%
  \BibitemOpen
  \bibfield  {author} {\bibinfo {author} {\bibfnamefont {J.~S.}\ \bibnamefont {Bell}},\ }\bibfield  {title} {\bibinfo {title} {On the einstein podolsky rosen paradox},\ }\href {https://doi.org/10.1103/PhysicsPhysiqueFizika.1.195} {\bibfield  {journal} {\bibinfo  {journal} {Physics}\ }\textbf {\bibinfo {volume} {1}},\ \bibinfo {pages} {195} (\bibinfo {year} {1964})}\BibitemShut {NoStop}%
\bibitem [{\citenamefont {Boldis}\ and\ \citenamefont {Lévay}(2024)}]{Boldis2024}%
  \BibitemOpen
  \bibfield  {author} {\bibinfo {author} {\bibfnamefont {B.}~\bibnamefont {Boldis}}\ and\ \bibinfo {author} {\bibfnamefont {P.}~\bibnamefont {Lévay}},\ }\bibfield  {title} {\bibinfo {title} {Segmented strings and holography},\ }\href {https://doi.org/10.1103/PhysRevD.109.046002} {\bibfield  {journal} {\bibinfo  {journal} {Physical Review D}\ }\textbf {\bibinfo {volume} {109}},\ \bibinfo {pages} {046002} (\bibinfo {year} {2024})}\BibitemShut {NoStop}%
\end{thebibliography}%

\end{document}